\documentclass[%
 aip,pof,
amsmath,
amssymb,
reprint %
]{revtex4-2}
\usepackage{nicefrac}
\usepackage{xcolor}
\usepackage{graphicx}
\usepackage{dcolumn}
\usepackage{enumitem}
\usepackage{newtxtext}
\usepackage[cmbraces,varbb,varvw]{newtxmath}
\usepackage{aecompl}
\usepackage{etoolbox}
\usepackage[scaled=0.95]{helvet}
\usepackage{titlesec}
\usepackage[colorlinks = true, linkcolor = blue, citecolor = blue, urlcolor = blue, linkbordercolor = white]{hyperref}
\usepackage{fancyhdr}
\titleformat{\section}
  {\normalfont\sffamily\large\bfseries\color{black}}
  {\thesection.}{1em}{}
\titleformat{\subsection}
  {\normalfont\sffamily\large\bfseries\color{black}}
  {\Alph{subsection}.}{1em}{}

\makeatletter
\def\@email#1#2{%
 \endgroup
 \patchcmd{\titleblock@produce}
  {\frontmatter@RRAPformat}
  {\frontmatter@RRAPformat{\produce@RRAP{*#1\href{mailto:#2}{#2}}}\frontmatter@RRAPformat}
  {}{}
}%

\makeatletter
\let\@msm@th@eqref\eqref
\renewcommand{\eqref}[1]{%
  \begingroup
 \leavevmode
  \color{blue}%
  \hypersetup{linkbordercolor=[named]{blue}}%
  \@msm@th@eqref{#1}%
  \endgroup
}
\makeatother

\newcommand*{\citen}{}
\DeclareRobustCommand*{\citen}[1]{%
  \begingroup
    \romannumeral-`\x 
    \setcitestyle{numbers}%
    \cite{#1}%
  \endgroup
}

\makeatother
\begin{document}

\preprint{AIP/123-QED}

\title[\centerline{\normalsize PHYSICS OF FLUIDS}]{\href{https://doi.org/10.1063/1.4887817}{{\textbf{Bag breakup of low viscosity drops in the presence of a continuous air jet}}}\\ \vspace{10pt}}

\author{V. Kulkarni}
\altaffiliation[Author to whom correspondence should be addressed. Electronic mail: ]{\href{mailto:varun14kul@gmail.com}{varun14kul@gmail.com}}
\author{P. E. Sojka}
\affiliation{Maurice J. Zucrow Laboratories, Purdue University, West Lafayette, Indiana 47906, USA \looseness=-1}


\received[Received ]{8 January 2014} \accepted[accepted ]{21 June 2014} \published[published online ]{16 July 2014}

\begin{abstract}
\vspace{10pt}
\centering
\begin{minipage}{\dimexpr\paperwidth-10.2cm}
This work examines the breakup of a single drop of various low viscosity fluids as it deforms in the presence of continuous horizontal air jet. Such a fragmentation typically occurs after the bulk liquid has disintegrated upon exiting the atomizer and is in the form of an ensemble of drops which undergo further breakup. The drop deformation and its eventual disintegration is important in evaluating the efficacy of a particular industrial process, be it combustion in automobile engines or pesticide spraying in agricultural applications. The interplay between competing influences of surface tension and aerodynamic disruptive forces is represented by the Weber number, \textit{We}, and Ohnesorge number, \textit{Oh}, and used to describe the breakup morphology. The breakup pattern considered in our study corresponds to that of a bag attached to a toroidal ring which occurs from $\sim 12< We < \sim 16$. We aim to address several issues connected with this breakup process and their dependence on \textit{We} and \textit{Oh} which have been hitherto unexplored. The \textit{We} boundary at which breakup begins is theoretically determined and the expression obtained, $\displaystyle We = 12(1 + \nicefrac{\normalsize 2}{\normalsize 3}Oh^2)$, is found to match well with experimental data [L.-P. Hsiang and G. M. Faeth, Int. J. Multiphase Flow 21(4), 545–560 (1995)] and [R. S. Brodkey, “Formation of drops and bubbles,” in \textit{The Phenomena of Fluid Motions} (Addison-Wesley, Reading, 1967)]. An exponential growth in the radial extent of the deformed drop and the streamline dimension of the bag is predicted by a theoretical model and confirmed by experimental findings. These quantities are observed to strongly depend on We. However, their dependence on \textit{Oh} is weak.
\end{minipage}
\end{abstract}
\maketitle


\section{\textbf{INTRODUCTION}\label{Intro}} 

Combustion of liquid fuels typically requires the fuel to be adequately atomized. The disintegration of bulk fluid into smaller fragments, thus, is of utmost significance \cite{Lefebvre1989}. To aid in process modeling, breakup is often thought of as a two-stage process consisting of primary and secondary atomization. Primary atomization involves the fragmentation of bulk liquid into smaller droplets. Secondary atomization on the other hand refers to further disintegration of these droplets due to external forces, such as the aerodynamic forces considered here. In practice, a plume of droplets interacts also with each other. Nevertheless, an understanding of the physics involving the interaction of these drops must begin with an analysis of breakup of a single drop.

Typically, three methods have been employed to study drop breakup of such a type. These are: Shock tube,\citep{Theofanous2011, Chou1998, Hsiang1995, Hsiang1992, Hsiang1993, Wierzba1988, Dai2001}, Drop tower \cite{Villermaux2009, Lane1951, Dodd1960}, and continuous jet\cite{Flock2012, Krzeczkowski1980, Hinze1955, Cao2007, Zhao2010, Zhao2011a, Zhao2011b, Zhao2011c, Opfer2014, Liu1997} techniques. Our investigation focuses on the continuous air jet method, i.e., the drop disintegration in the presence of an enveloping air jet. It is worthwhile to add here that the results of the continuous jet method and shock tube investigations are, in general, comparable as pointed by Ref. \citen{Guildenbecher2009} where a criteria based on droplet/continuous air jet velocity, physical properties, and characteristic times involved in the process, is established. However, there might be some deviations which are pointed out wherever required. 

The study of aerodynamic disintegration of a single drop owes its origin to the problem of determining the shape of a falling raindrop. Rain drops for a long time were believed to be tear shaped. The results of Ref. \citen{Lenard1904} showed that large drops were not tear shaped but flattened from one side in the vertical direction. This was attributed to internal circulations inside the drop which were set up due to tangential friction of air which brought the mass of the drop into relative motion. Subsequently, Refs. \citen{Dodd1960} and \citen{Flower1927} amongst others, confirmed and expanded on the ideas of Ref. \citen{Guildenbecher2009} which have been adequately summarized by Ref. \citen{McDonald1954}. 

As the air velocity around the liquid drop was increased additional breakup patterns\cite{Krzeczkowski1980, Hinze1955} quite different from the one mentioned above were observed. A convenient way to characterize these breakup morphologies is to quantify them in terms of regime boundaries based on characteristic non-dimensional groups. Such a dimensional analysis yields the following numbers:
\begin{equation}\label{Eqn1}
\displaystyle We = \dfrac{\rho_a U^2 d_0}{\sigma}
\end{equation}
\begin{equation}\label{Eqn2}
Oh = \dfrac{\mu_l}{\sqrt{\rho_l\sigma d_0}}
\end{equation}
\begin{equation}\label{Eqn3}
Re = \dfrac{\rho_a U d_0}{\mu_a}
\end{equation}
Here, $\rho_a$ is the density of air; $\rho_l$ is the density of the liquid; $\sigma$ is the surface tension; $d_0$ is the initial drop diameter; $\mu_a$ is the viscosity of the gas (in our case air); and $U$ is the mean air velocity. Based on the relative magnitude of aerodynamic and restoring surface tension forces given by the Weber number, $We$, the following breakup modes (corresponding to continuous jet breakup and $Oh$ < 0.1) are observed\cite{Flock2012, Krzeczkowski1980, Hinze1955, Cao2007, Zhao2010, Zhao2011a, Zhao2011b}: (\textit{i})Vibrational ($We < \sim12$), (\textit{ii}) Bag Breakup ($\sim12 < We < \sim16$), (\textit{iii}) Bag-Stamen ($\sim16 < We < \sim30$), (\textit{iv}) Shear Breakup ($\sim30 < We < \sim41$). There is a strong dependence of these regime boundaries on $Oh$ for values of $Oh > 0.1$ which is depicted in the transitional $We$ plot given by Ref. \citen{Hsiang1995}. Of these various regimes our study focuses on bag breakup for two important reasons: (i) this regime marks the onset of guaranteed breakup of drops, a significant aspect from the standpoint of atomization and (ii) it is the precursor to multimode breakup\cite{Chou1998, Hsiang1995, Guildenbecher2009}. From the perspective of practical application, the $We$ at which bag breakup is observed is usually encountered in early injection strategies for homogeneous charge \cite{Chryssakis2008}. Also, bag breakup in the presence of a continuous air jet is responsible for distinct peaks in the drop size distributions of high flow rate industrial sprays\cite{Opfer2014}. The viscous nature of the fuel droplets motivates the need to include effects of $Oh$. It is also noteworthy to point out that bag breakup is seen in some other physical scenarios as well. For example, Ref. \citen{Cao2007} identified a new regime called the dual-bag breakup regime specific to the continuous air jet type of breakup and differs from the traditional bag-stamen breakup as the continuous air jet flow imposes a continuous body force which makes the $We$ of the core drop produced from the first bag breakup process large enough to form another bag. References \citen{Ng2008} and \citen{Scharfman2012} document cases of multiple bag formation where the liquid jet is disrupted by aerodynamics forces induced by the cross flowing air stream. Bag breakup of one drop, therefore, serves as an important canonical case for these studies. Thus, the aforementioned arguments underscore the importance of the problem both from the point of view of fundamental research as well as industrial applications. It also demonstrates the need to examine the effect of $We$ and $Oh$ within the bag breakup regime. Against this backdrop we shall delve deeper into various aspects of bag breakup and outline the precise objectives of our study along the way.

\begin{figure}[htp!]
   \centering
   \vspace{0pt}
   \includegraphics[scale=0.35]{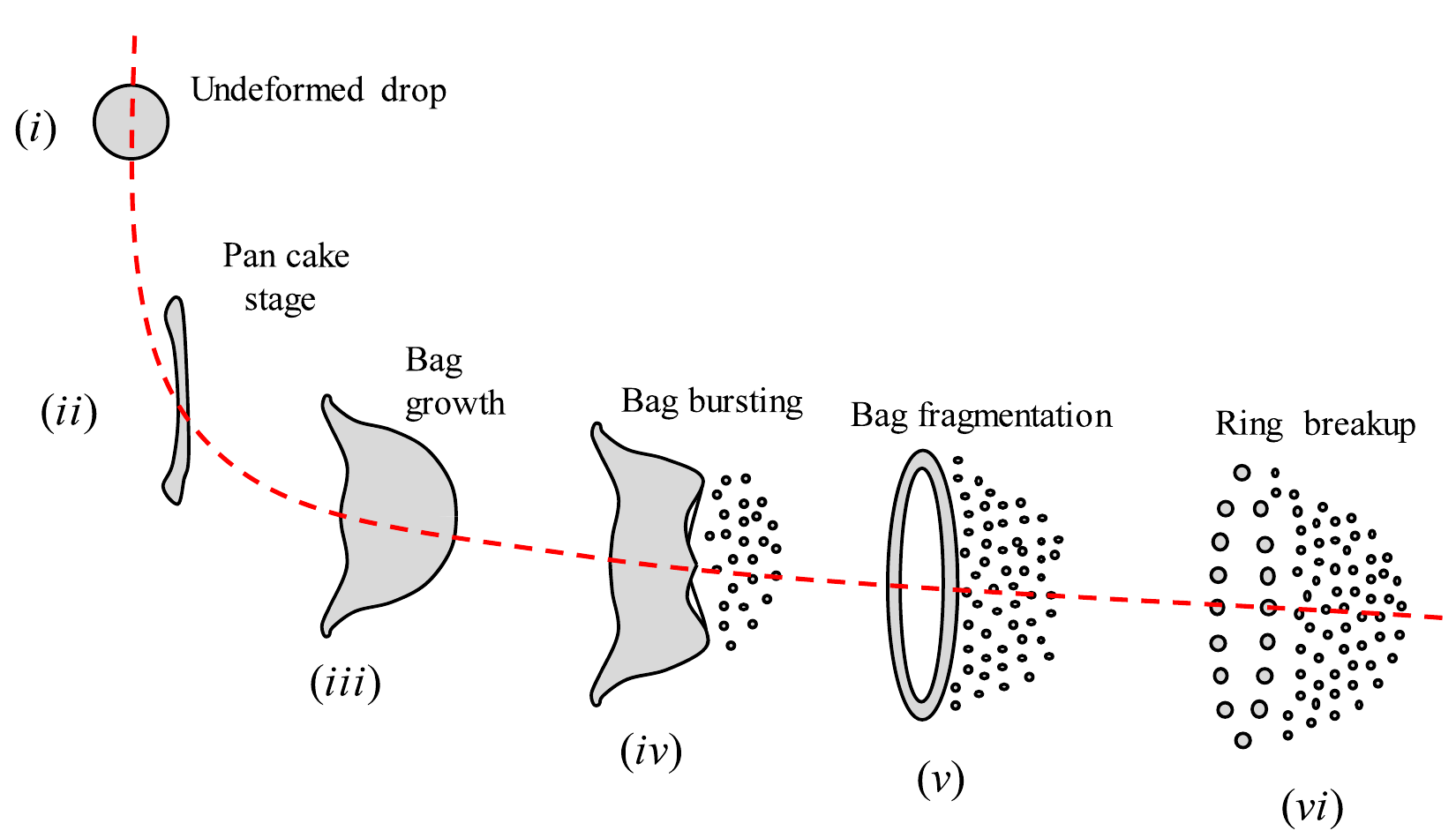}
   \vspace{0pt}
  \caption{Sketch showing deformation of drop as it moves through the air flow field.}
  \label{Fig1}
\end{figure}

The investigation of the process of bag breakup begins with the classification of the various stages of drop deformation as it transits from a spherical entity to an expanding bag which ultimately bursts. These topological changes are categorized into six commonly accepted stages\cite{Dodd1960, Hinze1955}.  Figure \ref{Fig1} shows an illustration of the same where the initially spherical drop is seen to deform to form a bag bounded by a thick rim before finally bursting.
\begin{enumerate}[leftmargin=*]
\setlength\itemsep{0.01em}
\item Initial drop of diameter, $d_0$.
\item Oblate spheroid (discoid) stage.
\item Bag growth.
\item Initiation of bag bursting.
\item Bag fragmentation.
\item Toroidal ring breakup.
\end{enumerate}

These topological changes are typically expressed in terms of radial and bag growth rate extents which are non-dimensionalised using the initial drop radius. Reference \citen{Chou1998} from their shock tube experiments give a detailed description of the bag temporal properties using shadowgraphs. Their findings showed a linear growth in the cross stream drop dimension/radial growth until the bag formation stage after which there is a parabolic variation ending with a linear trend towards the end when the toroidal ring breaks up. Similar investigation for falling raindrops using the drop tower technique by Ref. \citen{Villermaux2009} revealed an exponential variation at later times for the bag which is parabolic at the beginning of the bag expansion process. Reference \citen{Opfer2014} used the continuous jet technique (identical to the setup in this study except for the wind tunnel arrangement) for near inviscid drops and found the trend for bag growth and radial growth to be exponential as well. They also showed similarity of the phenomenon of aerodynamic drop deformation in the bag breakup regime to the corresponding phenomena of drop-wall impact and binary drop collision if the correct velocity scaling is used. In our efforts in this work we aim to experimentally and theoretically investigate these topological changes from the initial drop stage to the bag growth stage incorporating $We$ and $Oh$ corrections and, hence, extending their findings.

The next thing to consider is the regime limit delineation. Reference \citen{Hsiang1995} in a series of shock tube experiments provide a significant result in this direction which is summarized in the form of a regime map depicting these boundaries on a We versus $Oh$ plot. Interestingly, they also report that the oscillatory deformation or vibrational mode of breakup ceases to exist at $Oh \approx 0.3$ and bag breakup at $Oh \approx 4$. This leads to an important conclusion that these two types of breakups do not extend indefinitely on the $Oh$ axis, quite contrary to the other breakup modes. References \citen{Gelfand1996, Brodkey1967, Pilch1987} give useful experimental correlations to mark this breakup boundary in terms of $We$ and $Oh$. A comprehensive summary of the same can be found in Ref. \citen{Guildenbecher2009}. There has been some effort to theoretically predict the transition $We$ as seen in the studies of Refs. \citen{Cohen1994, Tarnogrodzki1993, Tarnogrodzki2001} . However, each of them suffers from some limitations which primarily revolve around being excessively empirical in nature. The recent work of Refs. \citen{Zhao2010} and \citen{Zhao2011a} considers aerodynamic loading of these drops and differs from the other studies \cite{Cohen1994, Tarnogrodzki1993, Tarnogrodzki2001}, in the sense that they posit a theory based on Rayleigh-Taylor instability and use the Rayleigh-Taylor wavenumber in the region of maximum cross stream dimension, \textit{N}\textsubscript{RT}, instead of $We$. Bag breakup, they claim would occur when \textit{N}\textsubscript{RT} is $\sqrt{13} - 1$ but this is restricted to only inviscid drops. In contrast to the above mentioned attempts, Ref. \citen{Villermaux2009} provides a fairly robust framework which is modified here to include the effects of viscosity and arrive at an expression which establishes this regime boundary between vibrational and bag breakup modes.

Although the experimental studies as mentioned thus far have been fairly exhaustive in their treatment of the problem of bag breakup relatively few computational studies have been carried out mainly because of the high costs involved to model the process accurately. References \citen{Jalaal2012, Jalaal2014, Poon2011} in some recent work have attempted to model this process numerically. Reference \citen{Jalaal2012} uses low cost, Direct Numerical Simulations (DNS) with an interface tracking mechanism to study bag breakup of drops. Their results are characterized in terms of the E\"{o}tv\"{o}s number, $Eo$, which is the ratio of buoyancy force and surface tension force and are 3D in nature which is a significant advancement from previous studies which were mainly 2D or axisymmetric. The bag breakup process is modelled starting from the initial drop deformation stage to the bag formation and evolution stage. The number of holes/punctures is reported to increase with $Eo$. Another scenario which has been numerically modelled in the broader context of drop deformation in uniform flow is the effect of transverse rotation of a droplet when released into a uniform flow. Reference \citen{Poon2011} uses the integral form of the Navier–Stokes equations which are solved in a discretized stream-function formulation using a finite volume staggered mesh method. The results are characterized in terms of the dimensionless rotation rate, $\Omega^*$ however, drop deformation before breakup is only considered. There is clearly a need for more numerical investigations which can validate all aspects of the bag breakup process especially the changes in topology. The current study aims to provide a foundation for more rigorous simulations in the future which shall help understand the bag breakup event holistically. 

To summarize, the review presented above has identified gaps in our existing knowledge of breakup of drops disrupted by aerodynamic forces. These include a theoretical and experimental investigation of (i) radial growth extent as a function of $We$ and $Oh$, (ii) bag growth extent as a function of $We$ and $Oh$, (iii) transition $We$ as a function of $Oh$ between the vibrational and bag breakup modes, (iv) initiation time $\left(T_{ini}\right)$ for secondary atomization process. In sections that follow we shall systematically investigate the highlighted issues starting with theoretical estimation of regime boundary followed by a quantification of radial and bag growth extents.

\section{EXPERIMENTAL SETUP} \label{ExpSetup}
\begin{figure}[htp!]
   \centering
   \vspace{0pt}
   \includegraphics[scale=0.40]{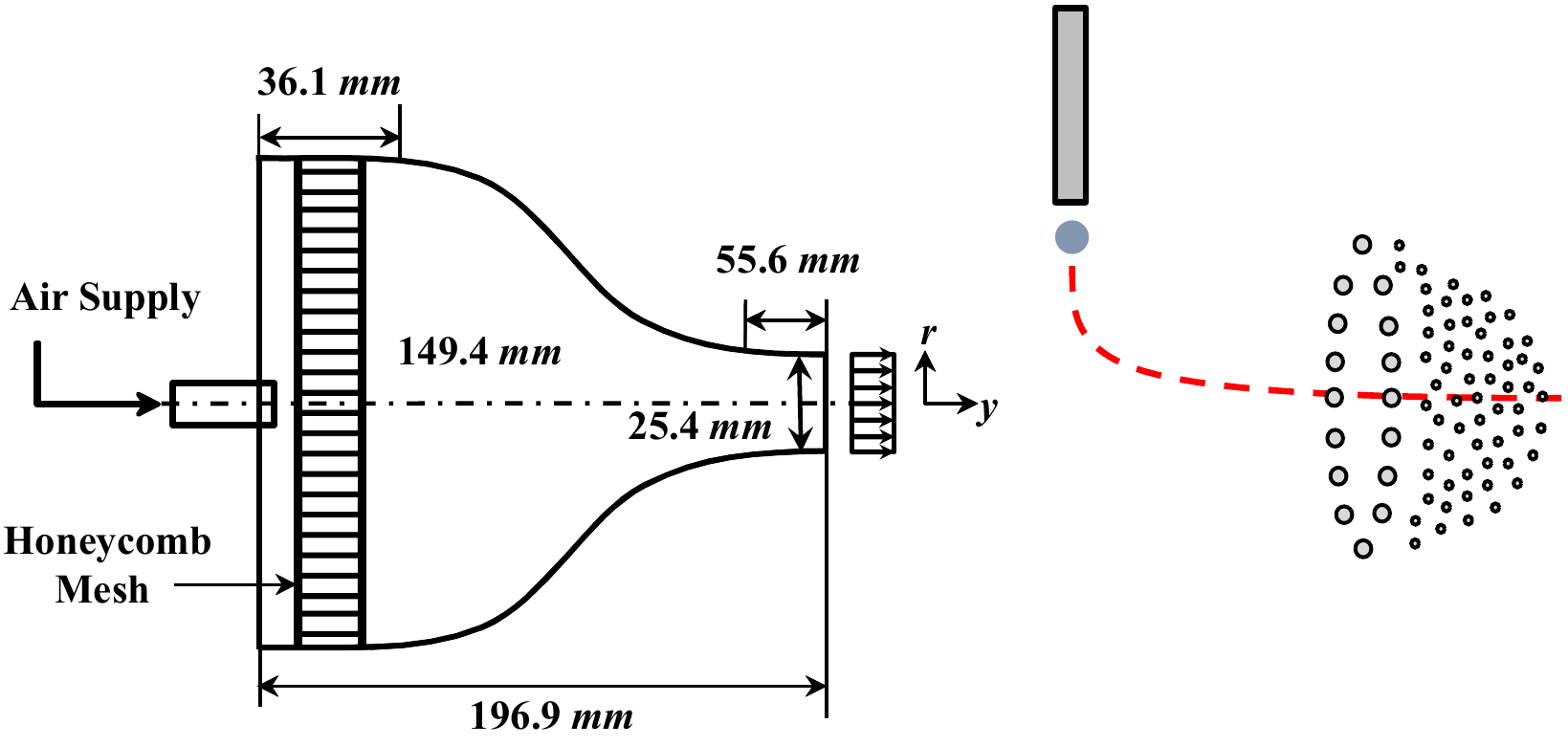}
   \vspace{0pt}
  \caption{Experimental setup.}
  \label{Fig2}
\end{figure}

The experimental setup used to study the drop breakup process is as shown in Fig. \ref{Fig2}. It consists of a drop generator, air supply, high speed imaging system, and backlight illumination. Drops are generated through a needle whose inner diameter is 0.25 mm. The pendant drops thus produced are roughly 10 times the needle inner diameter, measuring $\sim$2.5 mm in diameter. The exact diameter depends on the surface tension and viscosity of the solution used. These details are given in Table \ref{Table1}, where experimental uncertainty is estimated based on three repeated trials. The drops upon emanating from the needle deform gradually eventually breaking up as they are disrupted by a horizontal air jet generated by a nozzle. This is situated 25 cm downstream of the needle tip. The air nozzle is made from clear acrylic with a converging outlet section 2.54 cm in diameter designed to produce a near plug velocity profile at its exit. The honeycomb installed ensures a steady, laminar velocity profile. Other details of the air flow structure and nozzle design can be found in Ref. \citen{Flock2012}.
\begin{table}[htp!] \label{Table1}
\caption{\label{Table1}Solutions tested (fluid properties at 293 K, 1 atm). The viscosity has been measured using a rotational viscometer and surface tension using Wilhelmy plate method. The values are compared with the literature, D.I. Water: Ref. \citen{Wu2004}, Air: Ref. \citen{Batchelor1976} , Ethanol: Refs. \citen{Fox1998} and Ref. \citen{Rodriguez1994} Glycerine (40\%, 50\%, 60\%, 73\%): Ref. \citen{Glycerine1963}}
\begin{ruledtabular}
\begin{tabular}{ l c c c }
\textrm{Solution (\% weight)}& \textrm{Density, $\rho_l$ ($kg/m^3$)} & \textrm{Surface Tension, $\sigma$ (N/m)}& \textrm{Dynamic Viscosity, $\mu_l \times 10^4 (Pa-s)$ }\\
\colrule
D.I. Water          & 997   $\pm$ 2   & 0.0710 $\pm$ 0.008  &  8.94 $\pm$ 1 \\
Ethanol 		          & 800   $\pm$ 4   & 0.0240 $\pm$ 0.005  & 16 $\pm$ 2\\
40\% Glycerol 	  & 1100 $\pm$ 8   & 0.0662 $\pm$ 0.003  &  30 $\pm$ 2\\
50\% Glycerol 	  & 1130 $\pm$ 7   & 0.0651 $\pm$ 0.007  &  72 $\pm$ 4\\
63\% Glycerol 	  & 1162 $\pm$ 10 & 0.0648 $\pm$ 0.003  &  108 $\pm$ 6\\
70\% Glycerol 	  & 1185 $\pm$ 4   & 0.0640 $\pm$ 0.010  &  356 $\pm$ 9\\
\end{tabular}
\end{ruledtabular}
\end{table}

\subsection{High Speed Imaging \label{HiSpIm}} 
Vision Research Phantom v7 high speed digital camera is used to capture the videos. The images were recorded at 4700 fps with an exposure time of 100 $\mu s$ and resolution of 800 $\times$ 600 pixels. It consists of a 105 mm focal length lens (Nikon AF Micro Nikkor) attached to the camera body. The camera was placed perpendicular to the breakup plane and focused on the jet centerline. A 1000 W Xenon arc lamp (Kratos model LH151N) is used as a light source. 

\begin{figure}[htp!]
   \centering
   \vspace{0pt}
   \includegraphics[scale=0.30]{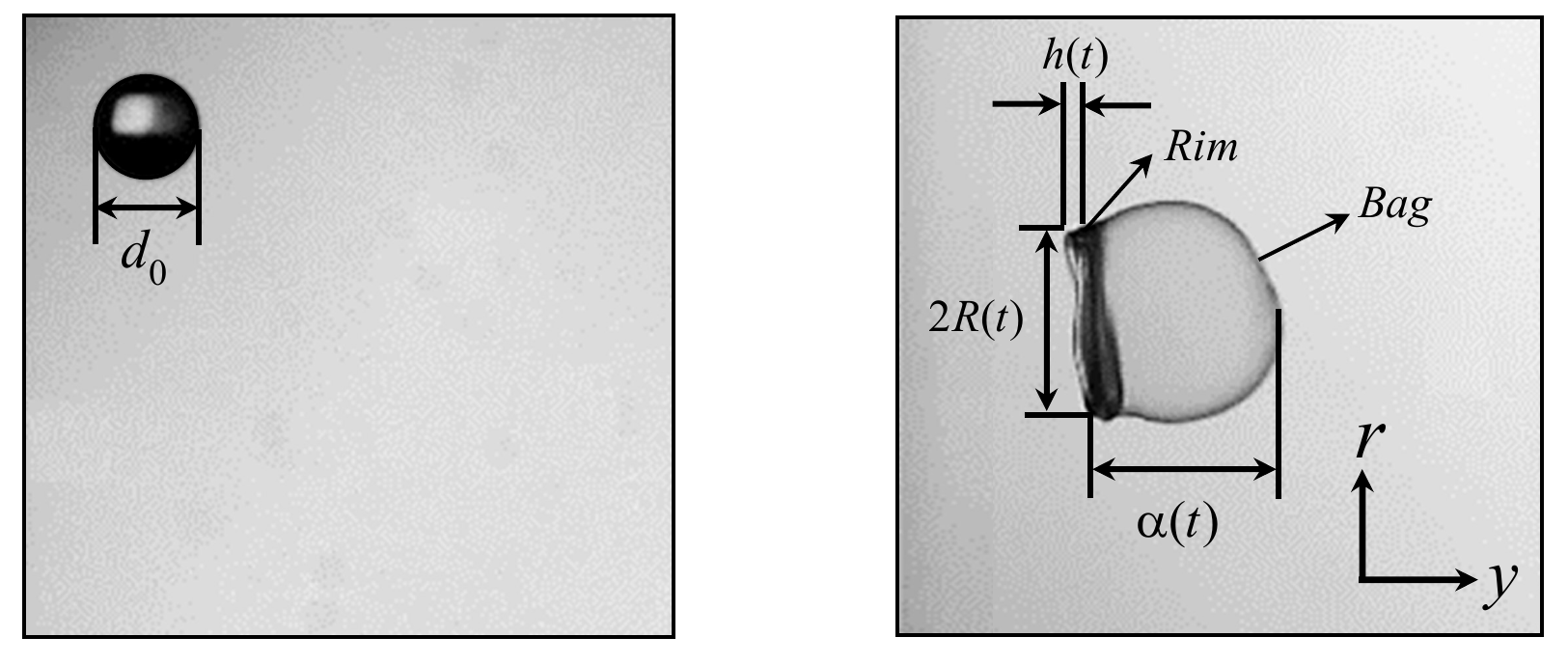}
   \vspace{0pt}
  \caption{Various parameters studied.}
  \label{Fig3}
\end{figure}

The measurements are made using the open source software IMAGE J (NIH) and Cine Viewer 663. Various parameters connected to droplet deformation are measured. Fig. \ref{Fig3} shows the various parameters quantified, which are: the thickness of the rim at any instant, $h\left(t\right)$, bag thickness, $\alpha\left(t\right)$, and the total radial extent $2R\left(t\right)$. 

\begin{figure}[htp!]
   \centering
   \vspace{0pt}
   \includegraphics[scale=0.35]{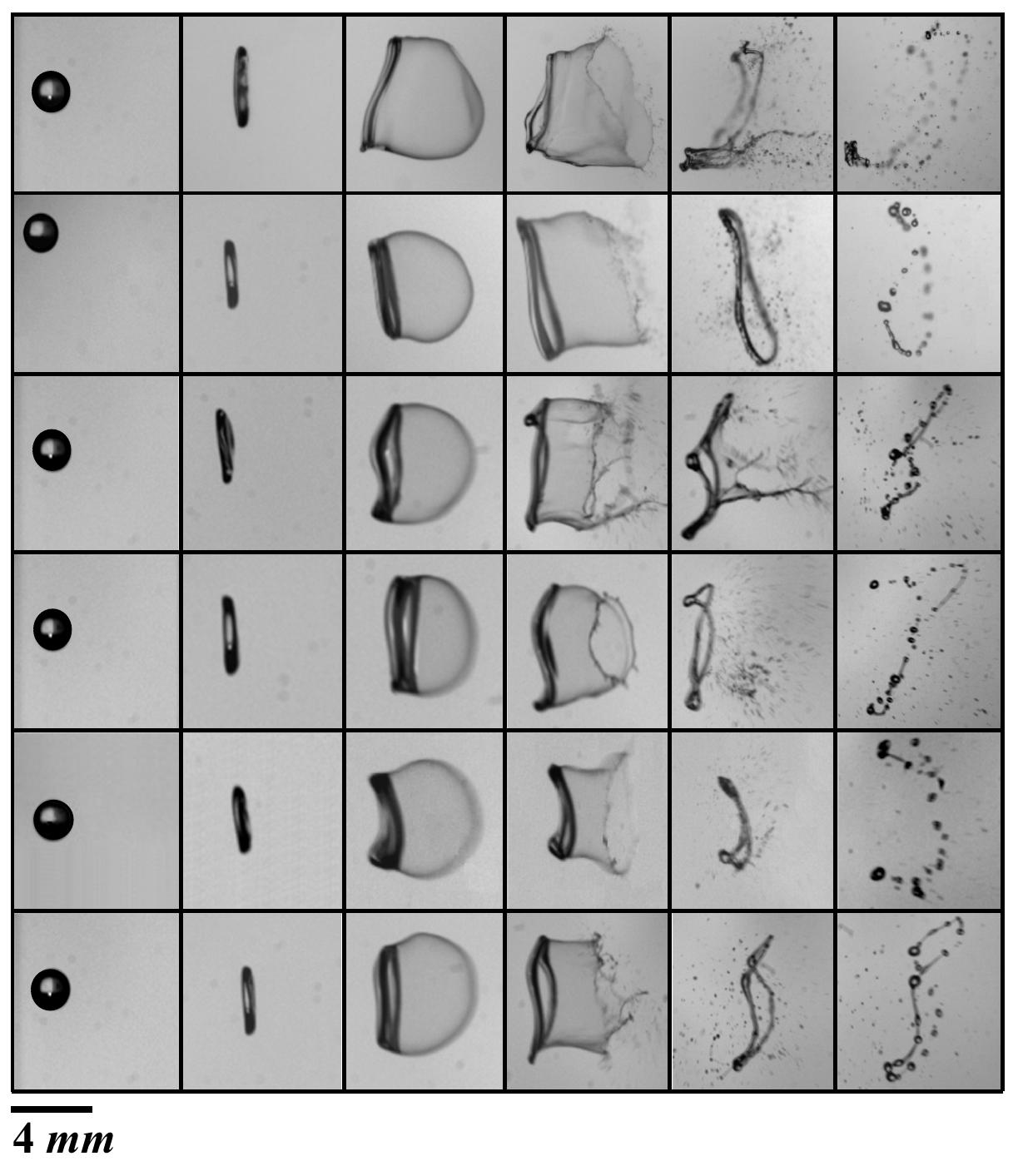}
   \vspace{0pt}
  \caption{Images showing bag breakup of drops. Top to bottom $Oh = 0.002, 0.007, 0.010, 0.015, 0.034, 0.058$ for $We \sim 14$. $\Delta t$ between successive frames is roughly 2 ms.}
  \label{Fig4}
\end{figure}

Three runs for each $We$ corresponding to each \textit{Oh} were taken and the values represent the arithmetic mean of measurements made from these videos. Bag breakup as captured for various \textit{Oh} and $We \sim 14$ is shown in Fig. \ref{Fig4} showing the different stages mentioned in Sec. \ref{Intro}.

\subsection{Test conditions \label{TestCond}} 
The different solutions tested are as listed in Table \ref{Table2}. USP grade glycerine and USP grade Ethyl alcohol is used. The property values are taken at 293 K and 1 atm. The glycerine solutions are prepared by mixing glycerine with deionized (D.I.) water to obtain various concentrations. The extent to which glycerine concentration in D.I. water could be increased was limited by the fact that drops with higher viscosity traverse larger distances axially and breakup at larger distances. Hence, the viscosity of solution is chosen such that the drops breakup within a distance of 1750 mm and the fragmentation process captured completely within the frame of the high speed camera. The experimental runs correspond to $We$ in the bag breakup regime, i.e., from 10 to 20 for inviscid and viscous drops. Table \ref{Table2} lists the conditions tested in terms of the non-dimensional parameters.

\begin{table}[htp!]
\caption{\label{Table2}Test conditions for experimental runs}
\begin{ruledtabular}
\begin{tabular}{ l c c c c c}
\multicolumn{6}{c}{Test conditions (at 293 K, 1 atm)} \tabularnewline
\colrule
\textrm{Solution (\% weight)}& \textrm{Drop diameter, $d_0$ ($mm$)} & \textrm{$We$}& \textrm{$Oh$}& \textrm{$Re$}& \textrm{Density ratio, $\rho_l/\rho_a$}\\
\colrule
D.I. Water          & 2.6 $\pm$ 0.10 & 14–16  & 0.002 & . . .  & 828\\
Ethanol 		          & 1.9 $\pm$ 0.10 & 12–14  & 0.007 & . . .  & 655\\
40\% Glycerol 	  & 2.1 $\pm$ 0.10 & 13–14  & 0.010 & . . .  & 912\\
50\% Glycerol 	  & 2.3 $\pm$ 0.10 & 12–15  & 0.015 & . . .  & 934\\
63\% Glycerol 	  & 2.0 $\pm$ 0.10 & 12–15  & 0.034 & . . .  & 964\\
70\% Glycerol 	  & 2.2 $\pm$ 0.10 & 12–15  & 0.058 & . . .  & 980\\
Air      			          &          . . .          &   . . .     &   . . .   & 1320–2820 & 1\\
\end{tabular}
\end{ruledtabular}
\end{table}

\section{DROP DEFORMATION DYNAMICS \label{DDDyn}} 
Drop deformation happens as soon as the drop enters the periphery of the horizontal air jet. The uneven pressure distribution results in changes in drop shape. The ensuing mathematical modelling is for the inviscid case which we extend to the viscous case. Hence, the air flow field calculations remain the same as those in Ref. \citen{Villermaux2009} but the liquid flow field analyses is modified accounting for viscous corrections. Coordinate directions are such that the cross-stream direction is radial,$r$, and the streamwise direction is axial, $y$. The model presented is one-dimensional.

\subsection{Air flow field \label{Airflowfield}} 
We begin with the air flow field description provided by Ref. \citen{Villermaux2009}, parts of which are repeated here for the sake of completeness. Let $U_r$ and $U_y$ be the velocities of air in the radial and streamwise directions. At this stage we are modeling between stages (i) and (ii) in Fig. \ref{Fig1}. Following Ref. \citen{Batchelor1976}, we assume the local air flow has the structure of a stagnation point, $U_y = - a\frac{U_y}{d_0}$. The value of $a$ is an indicator of the rate of stretching and $U$ is the mean velocity of the air jet as it exits the nozzle. 

In the inviscid, incompressible, quasi-steady assumption conservation of air momentum and mass can be solved to give a value of the pressure field around the drop\footnote{The coefficient of the, $y^2$ term has been reported as $\frac{a^2U^2}{8d_0^2}$ whereas it should be $\frac{a^2U^2}{2d_0^2}$ as written here.} as which as  $p_a\left(r,y\right) = p_a\left(0\right) - \rho_a \frac{a^2U^2}{8d_0^2} r^2 + \rho_a \frac{a^2U^2}{2d_0^2} y^2$ which at $y = 0$ leads to,
\begin{equation}\label{Eqn4}
p_a\left(r\right) = p_a\left(0\right) - \rho_a \frac{a^2U^2}{8d_0^2} r^2
\end{equation}
where $p_a\left(r\right)$ is the air pressure and $p_a\left(0\right) = \frac{\rho_a U^2}{2}$ is the stagnation pressure at $r = y = 0$.
From the conservation of mass we also get $U_r$ from which we can compute the net air velocity as $U_{net} = \sqrt{U_r^2 + U_y^2}$. This value is subsequently used for comparison with PIV measurements of Ref. \citen{Flock2012} to determine the stretching factor, $a$.

\subsection{Liquid flow field \label{Liquidflowfield}}  
For the deforming liquid we solve the viscous Navier-Stokes equations in cylindrical coordinates: 
\begin{equation}\label{Eqn5}
\rho_l{\left(\dfrac{\partial u_r}{\partial t} + \dfrac{\partial u_r}{\partial r}\right)} = -\dfrac{\partial p_l}{\partial r} + \mu \left[\dfrac{1}{r}\dfrac{\partial}{\partial r}\left(r \dfrac{\partial u_r}{\partial r}\right) - \dfrac{u_r}{r^2}\right]
\end{equation}

\begin{equation}\label{Eqn6}
r \dfrac{\partial h}{\partial t} + u_r \dfrac{\partial }{\partial r}\left(r u_r h\right) = 0
\end{equation}

From the differential form \footnote{see appendix section \ref{apx1} for details of the derivation} of the mass conservation, \eqref{Eqn6} and global mass conservation (given below).
\begin{equation} \label{Eqn7}
h\left(t\right) = \dfrac{d_0^3}{6 R^2\left(t\right)}
\end{equation}
we get $u_r\left(r,t\right)$ given by \footnote{see appendix section \ref{apx2} for details of the derivation of $u_r\left(r,t\right)$},
\begin{equation}\label{Eqn8}
u_r\left(r, t\right) = \dfrac{r}{R}\left(\dfrac{dR}{dt}\right)
\end{equation}
We shall verify expression \eqref{Eqn8} in Sec. \ref{Discoid}. 

Substituting \eqref{Eqn8} in \eqref{Eqn5} we see that the viscous stress terms cancel. This points to the fact the viscous stresses must enter the problem through the boundary conditions. First, we consider the normal stress balance across a fluid interface written as,
\begin{equation}\label{Eqn9}
\sigma \kappa = \mathbf{T}_{rr,a} - \mathbf{T}_{rr,l}
\end{equation}
Here, $\mathbf{T}_{rr,l}$ and $\mathbf{T}_{rr,a}$ represent the normal stress components associated with the liquid and the surrounding air and are given by $-p_l\left(r\right) +  2\mu_l \left(\frac{\partial u_r}{\partial r}\right)$ and $-p_a\left(r\right)$, respectively, at a given radial location $r$. Specifying \eqref{Eqn9} at $r = R\left(t\right)$ yields,
\begin{equation}\label{Eqn10}
p_l\left(R\right) = p_a\left(R\right)  + \sigma\kappa + 2\mu_l\left(\dfrac{\partial u_r}{\partial r}\right)
\end{equation}
\begin{figure}[htp!]
   \centering
   \vspace{0pt}
   \includegraphics[scale=0.35]{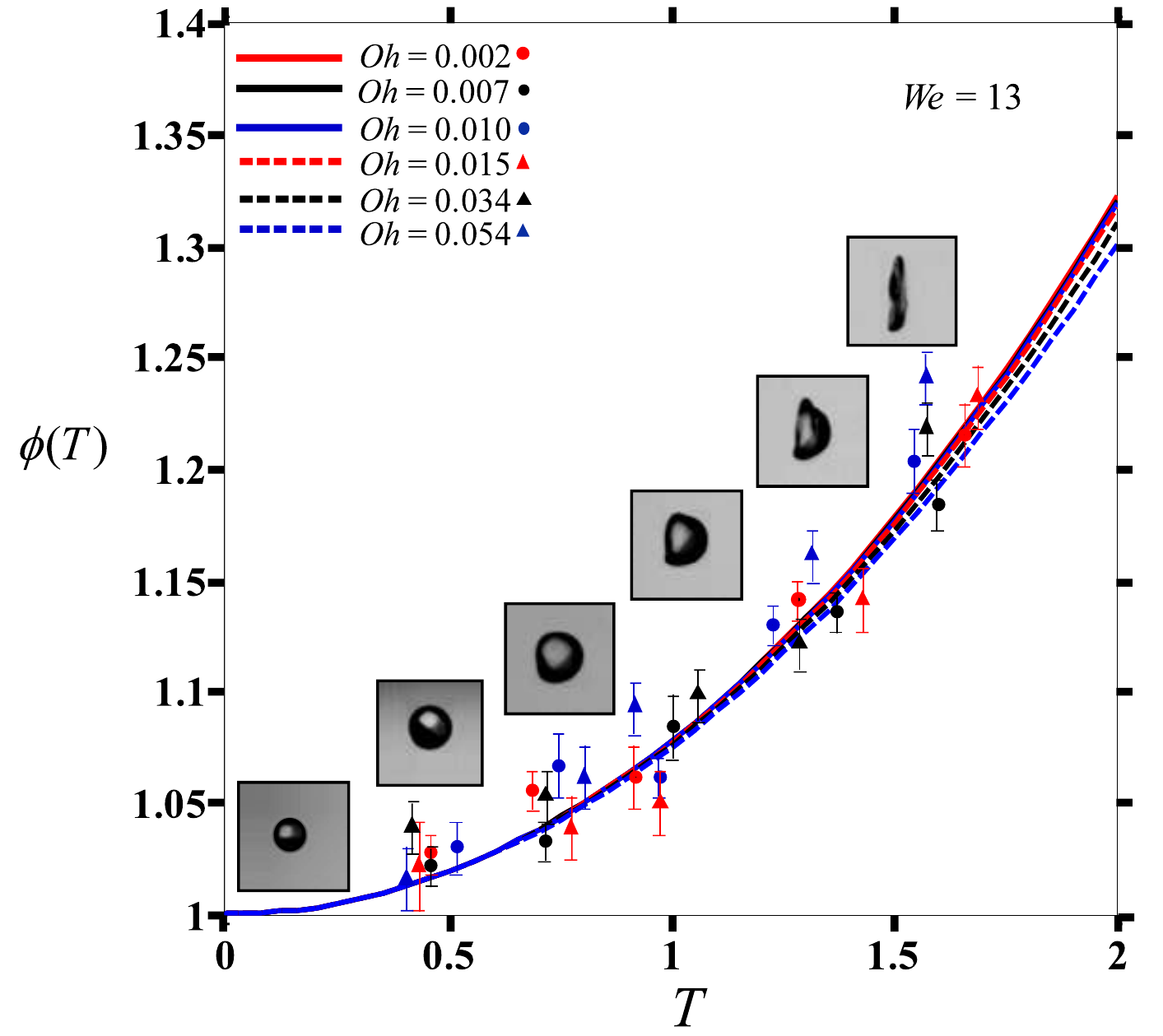}
   \vspace{0pt}
  \caption{Variation of $\phi\left(T\right)$ vs $T$ for a fixed $We ( = 13)$. Symbols are experimental data while lines represent theoretical results.}
  \label{Fig5}
\end{figure}
where $\sigma$ is the surface tension at the interface, $\kappa$ is the curvature of the interface at $r = R\left(t\right)$, and $\mu_l$ is the liquid dynamic viscosity. Using \eqref{Eqn4} in \eqref{Eqn10}, we get \footnote{In \textcolor{blue}{(}\ref{Eqn11}\textcolor{blue}{)} the $+$ sign in front of $\frac{2\mu_l}{R}\left(\frac{dR}{dt}\right)$ is mistakenly written as $-$ in the published version of the paper.}
\begin{equation}\label{Eqn11}
p_l\left(R\right) = p_a\left(0\right) - \rho_a \frac{a^2U^2}{8 d_0^2} R^2 + \dfrac{2\sigma}{h} + \dfrac{2\mu_l}{R}\left(\dfrac{dR}{dt}\right)
\end{equation}
In \eqref{Eqn10} $\kappa$ owing to the rounded periphery of the liquid disk evaluates to $\left(\frac{h\left(t\right)}{2}\right)^{-1}$. The second boundary condition, namely, the tangential stress balance is inconsequential as the stresses in that direction on the interface are negligible.

\subsection{Transition We and radial deformation $\boldsymbol{\phi}$(T)}

Integrating the momentum equation \eqref{Eqn5} between $r = 0$ and $r = R\left(t\right)$ taking advantage of \eqref{Eqn11} and non-dimensionalising the result thereafter using $\phi = R/\frac{d_0}{2}$ we obtain \footnote{\textcolor{blue}{(}\ref{Eqn5}\textcolor{blue}{)} in the published paper has a 8 as the prefactor to the second term which has been corrected to 16 here.},
\begin{equation}\label{Eqn12}
\frac{d^2 \phi}{dT^2}  + \frac{16Oh}{\sqrt{We}}\frac{1}{\phi^2}\frac{d \phi}{dT}  - 4\left(\dfrac{a^2}{16} - \dfrac{6}{We}\right) \phi = 0
\end{equation}
\begin{figure}[htp!]
   \centering
   \vspace{0pt}
   \includegraphics[scale=0.35]{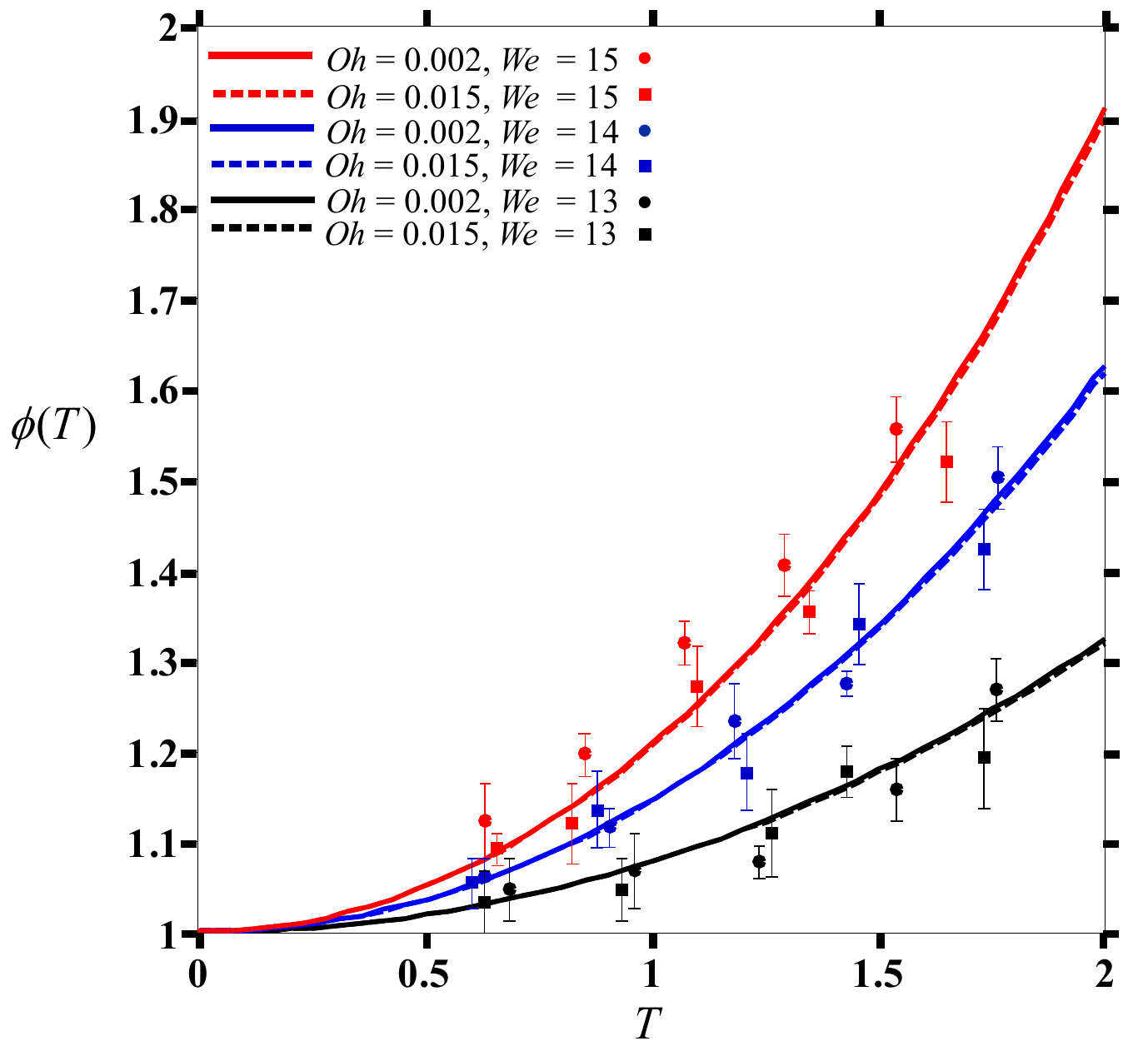}
   \vspace{0pt}
  \caption{Variation of $\phi\left(T\right)$ vs $T$ for varying $We ( = 13)$ and $Oh = 0.002, 0.015$. Symbols are experimental data while lines represent theoretical results.}
  \label{Fig6}
\end{figure}
$We$, $Oh$ are as defined in \eqref{Eqn1} and \eqref{Eqn2}. $\tau$ is the time scale typical to the process of secondary atomization and is given by,
\begin{equation}\label{Eqn13}
\tau = \dfrac{d_0}{U}\sqrt{\dfrac{\rho_l}{\rho_a}}
\end{equation}
$\tau^{-1}$ is particularly significant and denotes the frequency of oscillation when $Oh=0$. \eqref{Eqn12} is a second order nonlinear equation known as Lienard’s equation similar to a harmonic oscillator with damping and no closed form analytical solution exists for such an equation. The coefficient of $\frac{d\phi}{dT}$, $\left(\frac{16Oh}{We}\right)\phi^{-2}$ constitutes the nonlinear damping term which adds to significant damping even for small $Oh$. 

We can solve \eqref{Eqn12} numerically to analyze the variation of $\phi\left(T\right)$. The effect of $We$ for a given $Oh$ is seen to be significant as compared to the variation with $Oh$ for a given $We$ (Figs. \ref{Fig5} and \ref{Fig6}). 

We may also observe that the exponential trends exhibited are in agreement with earlier observations by Refs. \citen{Chou1998} and \citen{Zhao2010}. Also, Refs. \citen{Hsiang1995} and \citen{Zhao2010} have used the Rayleigh-Taylor instability mechanism to explain the ballooning of the oblate spheroidal structure to a bulging bag. It is interesting to note that such an explanation assumes an exponential ansatz pointing to the conclusion that the bag growth must be exponential, qualitatively, which is in agreement with our findings here.

\eqref{Eqn12} as such cannot be solved analytically however, we observe that from physical considerations the value of $\phi\left(T\right)$ must oscillate about the non-dimensional equilibrium radius. It must be noted here that the analysis in Ref. \citen{Villermaux2009} which may lead to the incorrect deduction that the drop oscillates about a mean corresponding to a zero drop radius can be understood more clearly in view of the discussion that follows. 

In view of the above arguments we seek an expansion for $\phi\left(T\right)$ as shown below\cite{Plesset1977}:
\begin{equation}\label{Eqn14}
\phi\left(T\right) \sim \phi_0\left(T\right) + \epsilon \phi_1\left(T\right) + O\left(\epsilon^2\right) \textrm{where,} \; |\epsilon| < 1.
\end{equation}
$\phi_0\left(T\right)$ is the equilibrium state of the drop and the subsequent terms are higher order corrections in $\epsilon$ to this nondimensional radius. Employing \eqref{Eqn14} in \eqref{Eqn12} we obtain,
\begin{equation}\label{Eqn15}
O\left(\epsilon^0\right):  \frac{d^2 \phi_0}{dT^2}  + \frac{16Oh}{\sqrt{We}}\frac{1}{\phi_0^2}\frac{d \phi}{dT}  - 4\left(\dfrac{a^2}{16} - \dfrac{6}{We}\right) \phi_0 = 0
\end{equation}
\begin{equation}\label{Eqn16}
O\left(\epsilon^1\right): \frac{d^2 \phi_0}{dT^2}  + \frac{16Oh}{\sqrt{We}}\left(\frac{1}{\phi_0^2}\frac{d \phi_1}{dT} - 2\frac{\phi_1}{\phi_0^3}\frac{d \phi_1}{dT}\right) - 4\left(\dfrac{a^2}{16} - \dfrac{6}{We}\right) \phi_1 = 0
\end{equation}
In \eqref{Eqn14} $O\left(\epsilon^2\right)$ terms are neglected, $O\left(\epsilon^0\right)$ terms correspond to the equilibrium position of the drop, and $O\left(\epsilon^1\right)$ terms represent the deviations about this equilibrium state. The balance between the surface tension and aerodynamics disrupting forces exists in the equilibrium state and the resulting expression for the diameter of the drop is,
\begin{equation}\label{Eqn17}
d_{eq} \sim \dfrac{\sigma}{\rho_a U^2}
\end{equation}
\eqref{Eqn16} now transforms into a linear second order differential and using the simplifications, $\frac{d\phi_1}{dT}\phi_0^{-2} >> 2\frac{\phi_1}{\phi_0^3}\frac{d\phi_0}{dT}$ and $\phi_0\left(T\right) \approx 1$, we arrive at the more familiar equation for a harmonic oscillator with damping:
\begin{equation}\label{Eqn18}
\frac{d^2 \phi_1}{dT^2} + \frac{16Oh}{\sqrt{We}}  \frac{d \phi_1}{dT} + |k_r| \phi_1 = 0
\end{equation}

\begin{figure}[htp!]
   \centering
   \vspace{0pt}
   \includegraphics[scale=0.35]{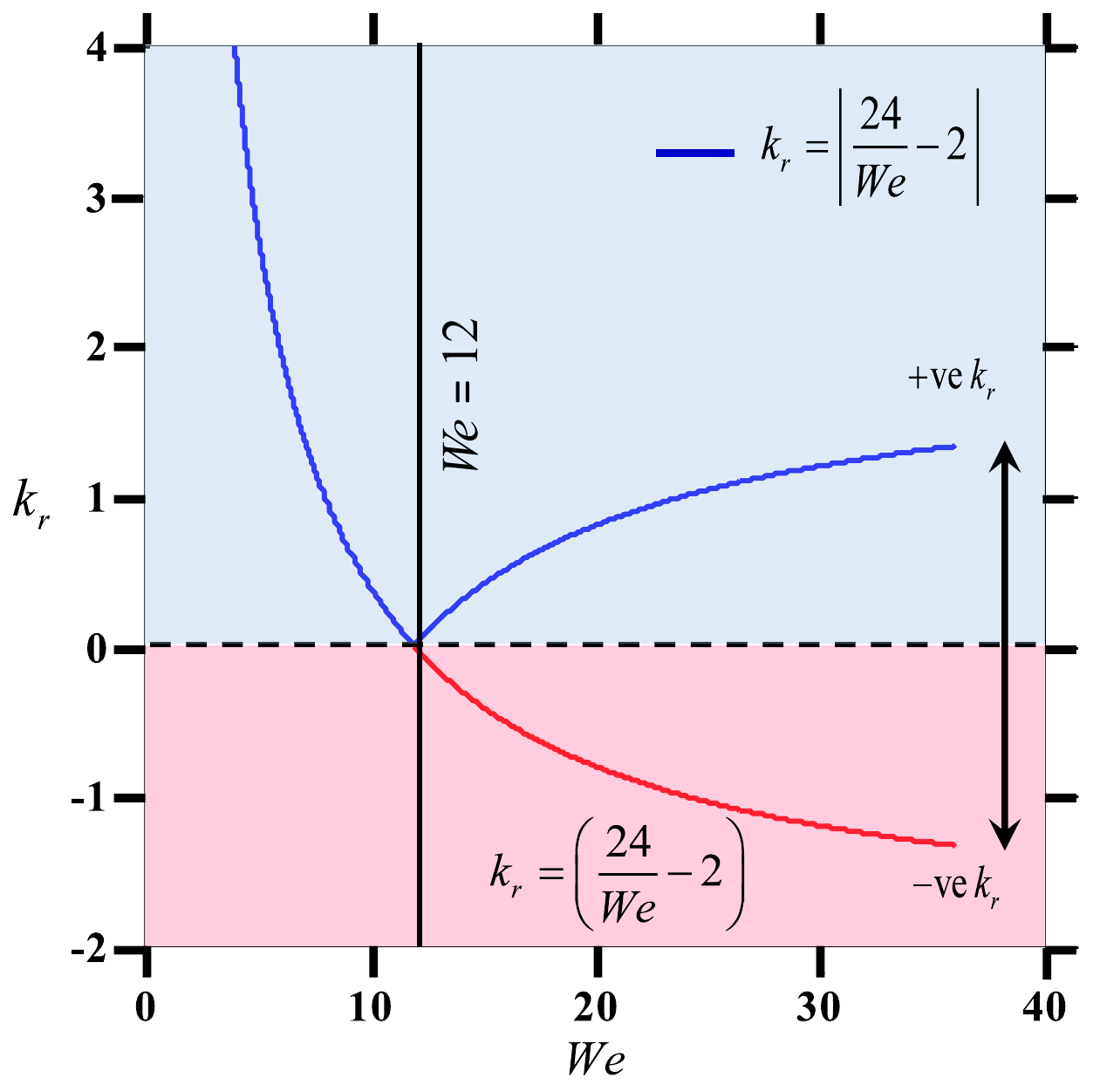}
   \vspace{0pt}
  \caption{Variation of $k_r$ with $We$.}
  \label{Fig7}
\end{figure}

The stiffness $k_r = k_r\left(a, We\right)$ is given by
\begin{equation}\label{Eqn19}
k_r = \left \{
     \begin{array}{ll}
      \quad\left(\dfrac{24}{We} - \dfrac{a^2}{4}\right)   \qquad We < \frac{96}{a^2}\\
      \quad\quad\quad\;\;0  \qquad\qquad                        We = \frac{96}{a^2}\\
     -\left(\dfrac{24}{We} - \dfrac{a^2}{4}\right)  \qquad\;We > \frac{96}{a^2}
    \end{array}
  \right.
\end{equation}
\eqref{Eqn18} is similar to that representing a linear spring mass damper system with the restoring force given by the coefficient of $\phi_1\left(T\right)$ and the damping term by the coefficient of $\frac{d\phi_1}{dT}$. The stiffness term \eqref{Eqn19} takes different values ranging from negative to positive depending upon the $We$ (Fig. \ref{Fig7}). In accordance with the nature of the restoring force we ensure that this value is positive which justifies the use of $|\;|$ with $k_r$. Intuitively, with increasing $Oh$ we expect the value of the critical $We$ number at transition to increase. Hence, to derive the condition for criticality we use $We > 96/a^2$ . Such a calculation considers that the drop oscillates even at $We > \frac{96}{a^2}$ which is true within $\frac{96}{a^2}\left(1 - \frac{2}{3}Oh^2\right)$ of $\frac{96}{a^2}$ . This seems to be an artifact \footnote{The resolution of this apparent inconsistency lies in assuming a varying value of the stretching factor, $a$ which is found to change with the shape of the deforming drop.} of the term $|k_r|$ of \eqref{Eqn18} which introduces the anomaly of growing solutions for $We > \frac{96}{a^2}$ for $Oh > 0$. As per Ref. \citen{Hsiang1995} oscillations in the drop cease beyond $Oh = 0.3$ which implies that this growth of $\phi_1\left(T\right)$ exists from $0.94\left(\frac{96}{a^2}\right)$ to $\left(\frac{96}{a^2}\right)$, i.e., 6 \% of $\left(\frac{96}{a^2}\right)$.

So, reverting to \eqref{Eqn18} and setting the condition for the solutions not to grow exponentially we obtain \eqref{Eqn20}:
\begin{equation}\label{Eqn20}
We_{cr} = We_{cr,\;Oh \to 0}\left(1 + \dfrac{2}{3}Oh^2\right)
\end{equation}
$We_{cr,Oh \to 0}$ is the critical $We$ at $Oh = 0$ or the inviscid case. In \eqref{Eqn20} for the limit of $Oh = 0$ we obtain $We_{cr,Oh \to 0} = \frac{96}{a^2}$.

We choose a value of a equal to $2\sqrt{2}$, whose validity is confirmed by the PIV measurements of Ref. \citen{Flock2012} which give an approximate value of $a$ as $3.0$ for ethanol drops. This leads to a value of $We_{cr,Oh \to 0}$ as 12 and is in good agreement with the value reported in reviews of Refs. \citen{Guildenbecher2009} and \citen{Pilch1987}. 

\begin{figure}[htp!]
   \centering
   \vspace{0pt}
   \includegraphics[scale=0.35]{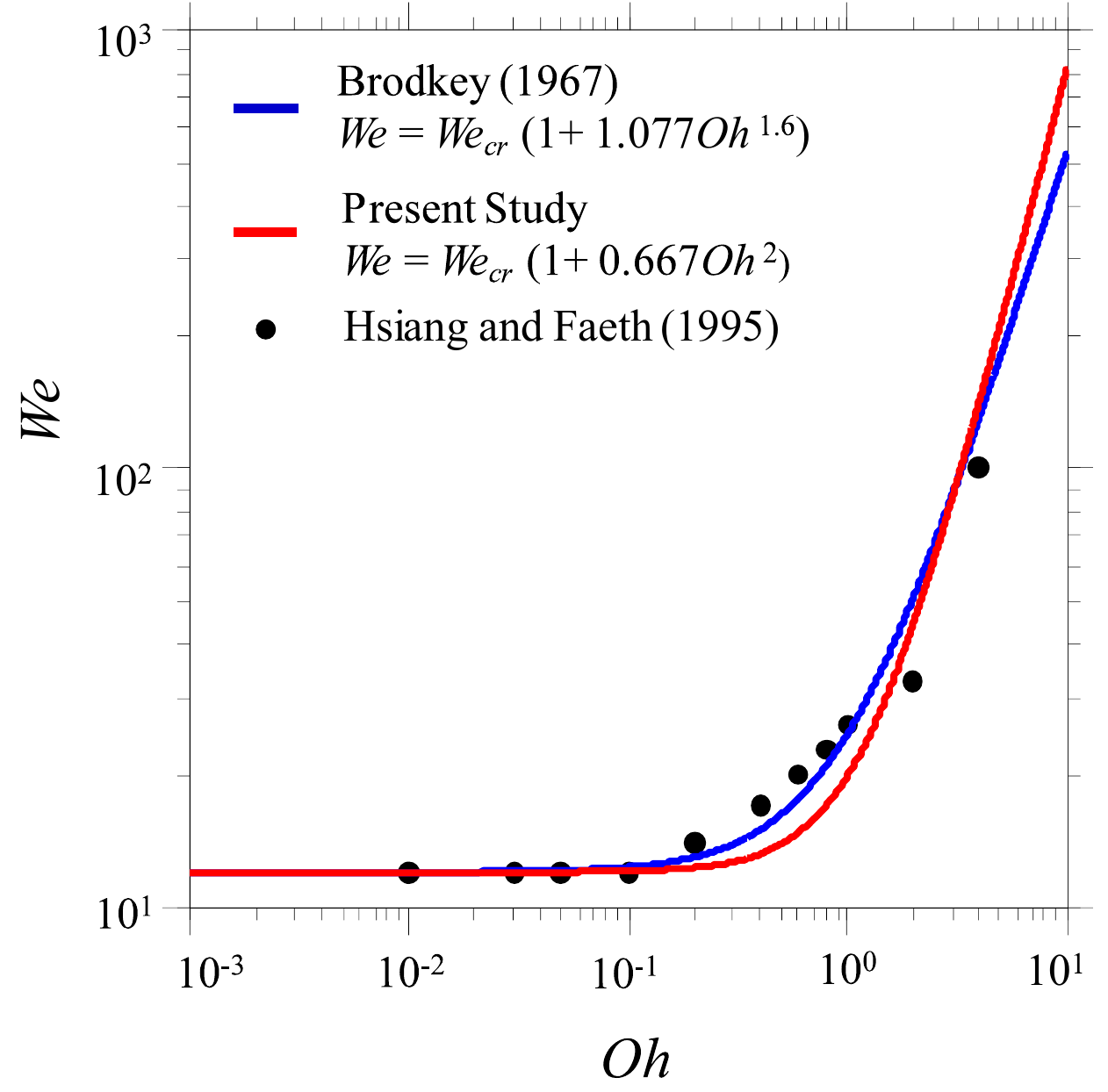}
   \vspace{0pt}
  \caption{Transition $We$ for bag breakup.}
  \label{Fig8}
\end{figure}

Figure \ref{Fig8} shows a comparison between \eqref{Eqn20}, experimentally observed values of Ref. \citen{Hsiang1995}, and empirically obtained correlation of Ref. \citen{Brodkey1967} for $Oh < 2.0$. 

To the authors’ knowledge this is the first theoretical relationship which does not involve use of any experimental correlations as seen in the works of Refs. \citen{Zhao2010, Zhao2011a} and \citen{Cohen1994, Tarnogrodzki1993, Tarnogrodzki2001}. Only value of the constant “\textit{a}” needs to be ascertained beforehand. Good match is found between the experimental and theoretical values validating expression \eqref{Eqn20}.

\eqref{Eqn18} also gives us information on the frequency of the damped vibrations associated with viscous drops. This value turns out to be $\sqrt{\frac{1}{\tau^2}\left(2 - \frac{24}{We}\right) - \frac{16Oh^2}{We}}$. Setting $Oh = 0$ we get for large $We$ the frequency of oscillations as $\tau^{-1}$ as expected for inviscid drops\cite{Villermaux2009}. $\tau$ is given by \eqref{Eqn13}.

It is also noteworthy that for $Oh = 0$ \eqref{Eqn15} and \eqref{Eqn16} before any simplification are decoupled, i.e., $\phi_0$ does not appear in \eqref{Eqn16}. Thus, the expression for the equilibrium diameter remains the same even for the inviscid case and \eqref{Eqn12} now reflects the oscillations of the inviscid drop about the equilibrium diameter as given by \eqref{Eqn17} . The expression for maximum radial extent in Ref. \citen{Villermaux2009} can be better understood in the light of the above arguments.

\subsection{Thickness of the oblate spheroid (discoid), H(T)} \label{Discoid}
\begin{figure}[htp!]
   \centering
   \vspace{0pt}
   \includegraphics[scale=0.35]{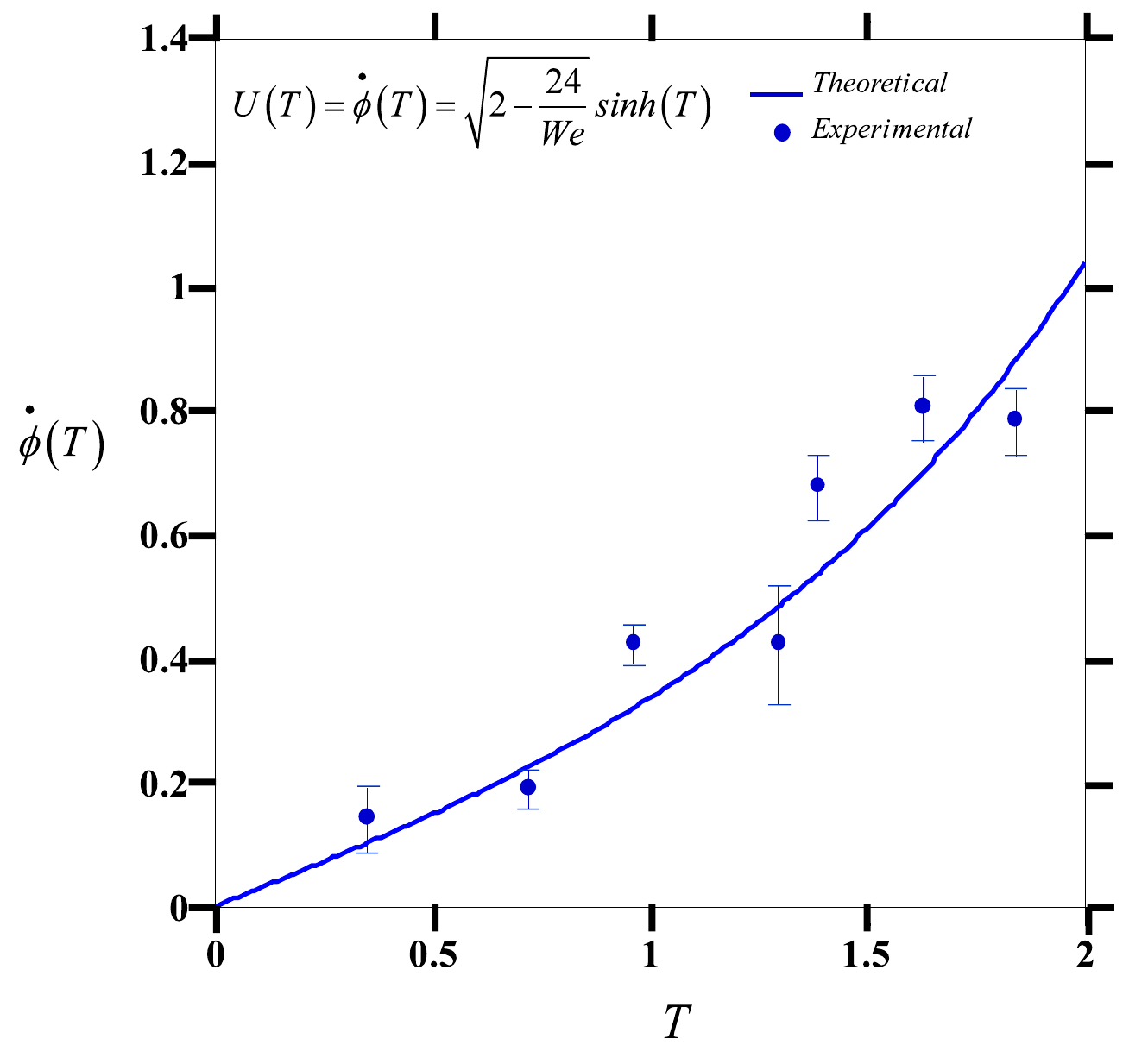}
   \vspace{0pt}
  \caption{Liquid flow field velocity, $U(T ) = u_r\left(r,t\right)|_{r=R}\dfrac{d_0}{\tau/2}$.}
  \label{Fig9}
\end{figure}

From \eqref{Eqn6} we obtain,
\begin{equation}\label{Eqn21}
\frac{dH}{dT} + \frac{2H}{\phi }\frac{d\phi}{dT} = 0
\end{equation}
In \eqref{Eqn21} $H\left(t\right) = \frac{h\left(t\right)}{d_0/2}$ which gives a value of $H\left(t\right)$ as,
\begin{equation}\label{Eqn22}
H \sim \phi^{-2}
\end{equation}
Solving \eqref{Eqn12} for the inviscid case, i.e., $Oh = 0$ we obtain the expression,
\begin{equation} \label{Eqn23}
\phi \sim e^{\sqrt{m}T}
\end{equation}
where $m = \left(2- \frac{24}{We}\right)$ takes the value of $a$ as $2\sqrt{2}$ and $\phi \sim cosh\left(\sqrt{m}T\right)$ in an exact sense but \eqref{Eqn23}  works well as an asymptotic solution for sufficiently large $T$. \eqref{Eqn23}  allows us to validate expression \eqref{Eqn8} for the velocity field inside the drop. Figure \ref{Fig9} shows the comparison between theory and experiment using the non-dimensional velocity $U\left(T\right)$ at $r = R$ for $Oh = 0.002$. Good agreement between the two is seen.

Another relevant parameter in the atomization context that we can extract from \eqref{Eqn23} is $T_{ini}$, the initiation time. Using \eqref{Eqn23} we compute the initiation time for inviscid drops as written below,
\begin{equation} \label{Eqn24}
T_{ini} \sim \dfrac{cosh^{-1} \phi_{max}}{\sqrt{2-\dfrac{24}{We}}}
\end{equation}

\eqref{Eqn24} depends on $\phi_{max}$ which is the value of $\phi$ just before bag formation. It is not possible to estimate
this from \eqref{Eqn12} as it predicts an indefinite growth of $\phi$ and does not predict when the bag formation will begin. One must note that $h\left(t\right)$ also does not shrink indefinitely in time as the rim eventually destabilizes. We, however, attempt to compare \eqref{Eqn24}  with existing correlations. Reference \citen{Zhao2011a} gives a value of $\phi_{max}$ as 2.15 for $20 < We < 80$ and $Oh < 0.1$. Substituting this in \eqref{Eqn24} $T_{ini}$ transforms to $We^{0.5}\left(We - 12\right)^{-0.5}$ giving a mean value of 1.34 which is somewhat close to an average value of 1.5 quoted in Ref. \citen{Guildenbecher2009}. However, in comparison to Ref. \citen{Pilch1987} correlation of $T_{ini} = 1.9 (We - 12)^{-0.25}$ the derived expression is an over prediction. This difference may be attributed to the shock tube results described in Ref. \citen{Pilch1987}  as against continuous jet technique used in our study.

\subsection{Bag growth, $\boldsymbol{\beta}\;$(T)}
The bag growth plays an important role in drop disintegration process. As the drop deforms and extends along cross stream direction while contracting in the streamwise direction there is unequal pressure distribution, $\Delta p$, around the drop. This process ends with gradual outward bulging of the oblate spheroidal structure which happens to accommodate the growing $\Delta p$ across the drop. In this section we attempt to model the bag shape and establish its relationship with $We$ and $Oh$. 

\begin{figure}[htp!]
   \centering
   \vspace{0pt}
   \includegraphics[scale=0.30]{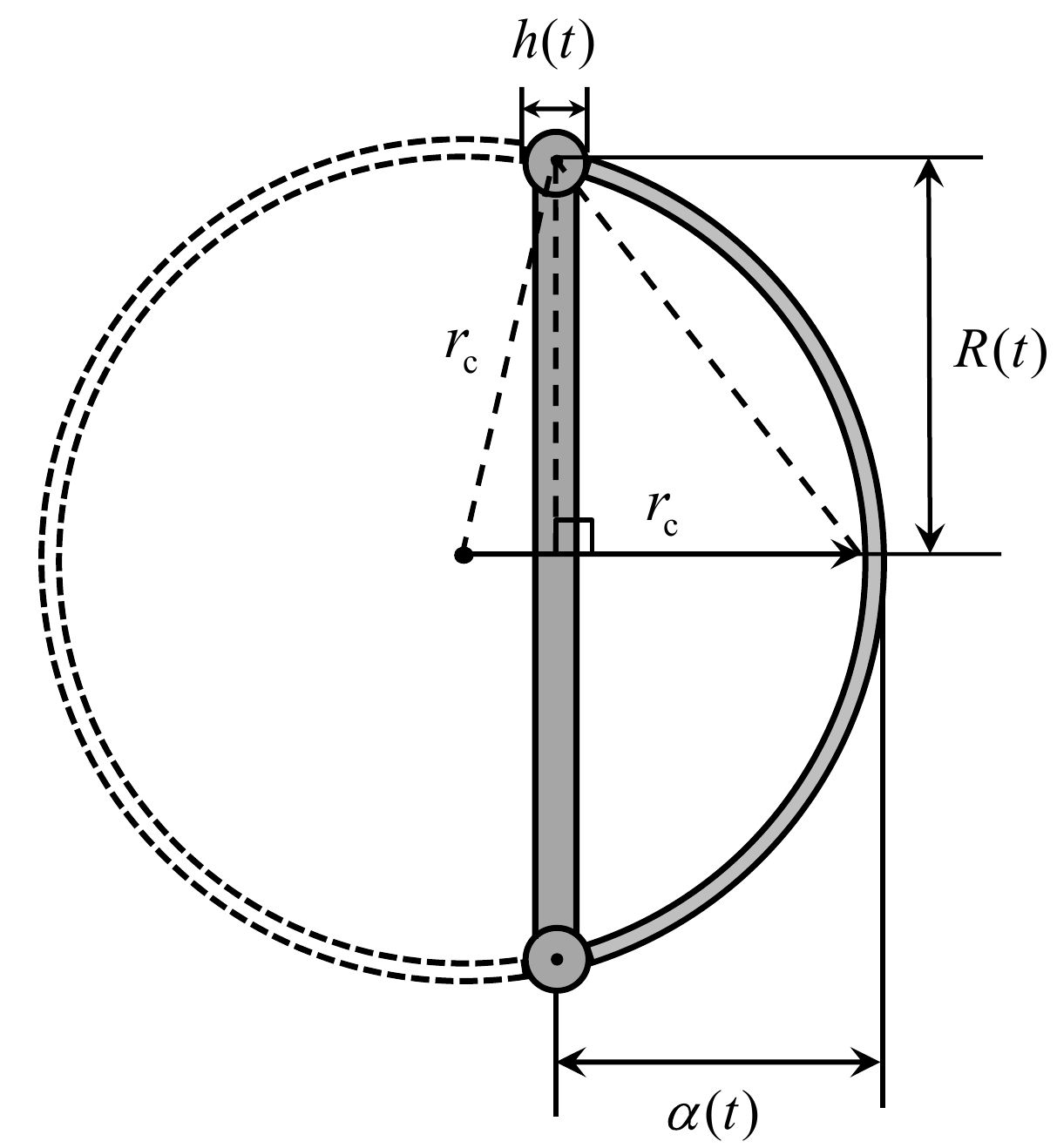}
   \vspace{0pt}
  \caption{Sketch of expanding bag with relevant quantities.}
  \label{Fig10}
\end{figure}

Writing the following force balance at the bag tip and making $We$ corrections to Ref. \citen{Villermaux2009} by incorporating the surface tension term, we obtain
\begin{equation}\label{Eqn25}
\dfrac{d^2 \beta}{dT^2} - \dfrac{48}{We}\beta - 2e^{T\sqrt{8 - \frac{96}{We}}} = 0
\end{equation}
Here, $\beta = \beta\left(T\right) = \alpha\left(T\right)/d_0/2$, where $\alpha\left(T\right)$ is shown in Figs. \ref{Fig3} and \ref{Fig10}. The initial conditions for this differential equation read as $\frac{d\beta}{dT}\left(0\right) = 0$ and $\beta\left(0\right) = 0$. The surface tension term involves evaluating the curvature term given by $4/r_c$ \footnote{The factor 4 accounts for the presence of two interfaces, much like a bubble} where $r_c$ varies with every point on the bag but since we are concerned with the dynamics near the bag tip this variation can be ignored. From geometrical considerations as shown in Fig. \ref{Fig10} we observe that the radius of curvature $r_c$ at the bag tip is,
\begin{equation} \label{Eqn26}
r_c = \dfrac{R^2\left(t\right) + \alpha^2\left(t\right)}{2\alpha\left(t\right)}
\end{equation}
Using \eqref{Eqn26} in the force balance at the bag tip we obtain \eqref{Eqn25}.
From \eqref{Eqn25} we observe bag growth which increases with increasing $We$ at a given instant of time $T$. It is worthwhile to note that \eqref{Eqn25} includes the effect of $We$ a fact not taken into consideration by Ref. \citen{Villermaux2009} even for the inviscid case. 

Employing viscous corrections in \eqref{Eqn25} and evaluating the viscous stress term $2\mu\frac{\partial u_{r_c}}{\partial r_c}$ with $u_{r_c} = \frac{d r_c}{dt}$ for the expanding bag (assuming spherical geometry) results in,
\begin{equation} \label{Eqn27}
\dfrac{d^2 \beta}{dT^2} - \dfrac{48}{We}\beta - \dfrac{12Oh}{\sqrt{We}}\beta^{-1}\dfrac{d\beta}{dT} - 2\phi^2 = 0
\end{equation}
Note that the same non-dimensionalization as used in \eqref{Eqn12} is invoked here. The initial conditions, $\beta(0) = 0$ may result in a singularity in \eqref{Eqn27} if $\beta^{-1}$ appears in \eqref{Eqn27} which is a consequence of evaluating $\frac{dr_c}{dt}$ using \eqref{Eqn26}. However, the two terms upon differentiation of \eqref{Eqn27} cancel out for very small $T$ at finite $\beta\left(T\right)$. We readily observe that \eqref{Eqn27} requires the value of $\phi\left(T\right)$ which can be obtained from \eqref{Eqn23} but that corresponds to the case of an inviscid drop \footnote{For our test conditions viscosity does not play a major role in determining the $\phi\left(T\right)$ so the expression for the inviscid case should work well (see Figs. \ref{Fig5}, \ref{Fig6}).}. 
\begin{figure}[htp!]
   \centering
   \vspace{0pt}
   \includegraphics[scale=0.30]{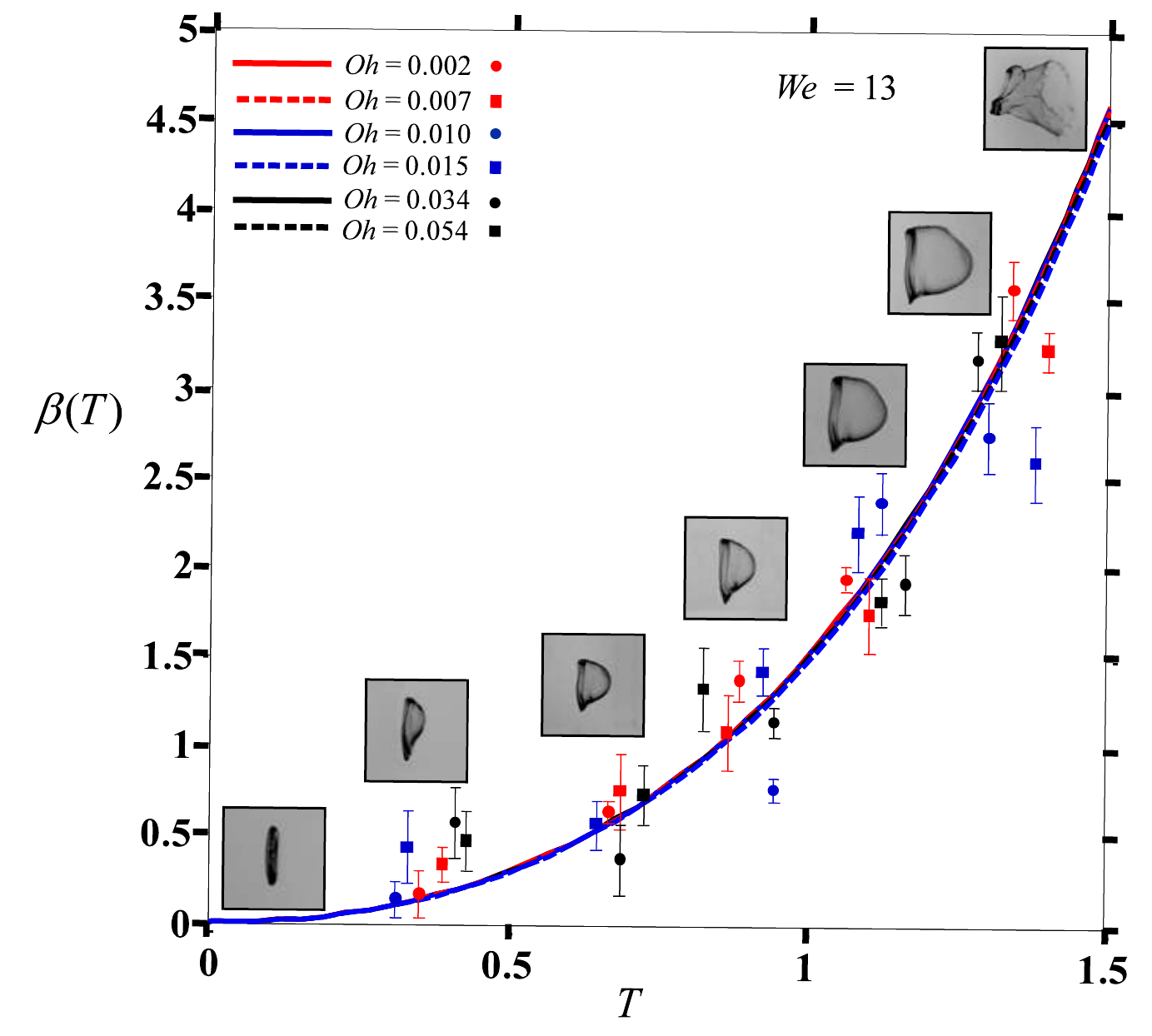}
   \vspace{0pt}
  \caption{Variation of $\beta\left(T\right)$ vs $T$ for varying $Oh$ and given $We$. Symbols with error bars represent experimental data while lines represent theoretical results.}
  \label{Fig11}
\end{figure}

\begin{figure}[htp!]
   \centering
   \vspace{0pt}
   \includegraphics[scale=0.30]{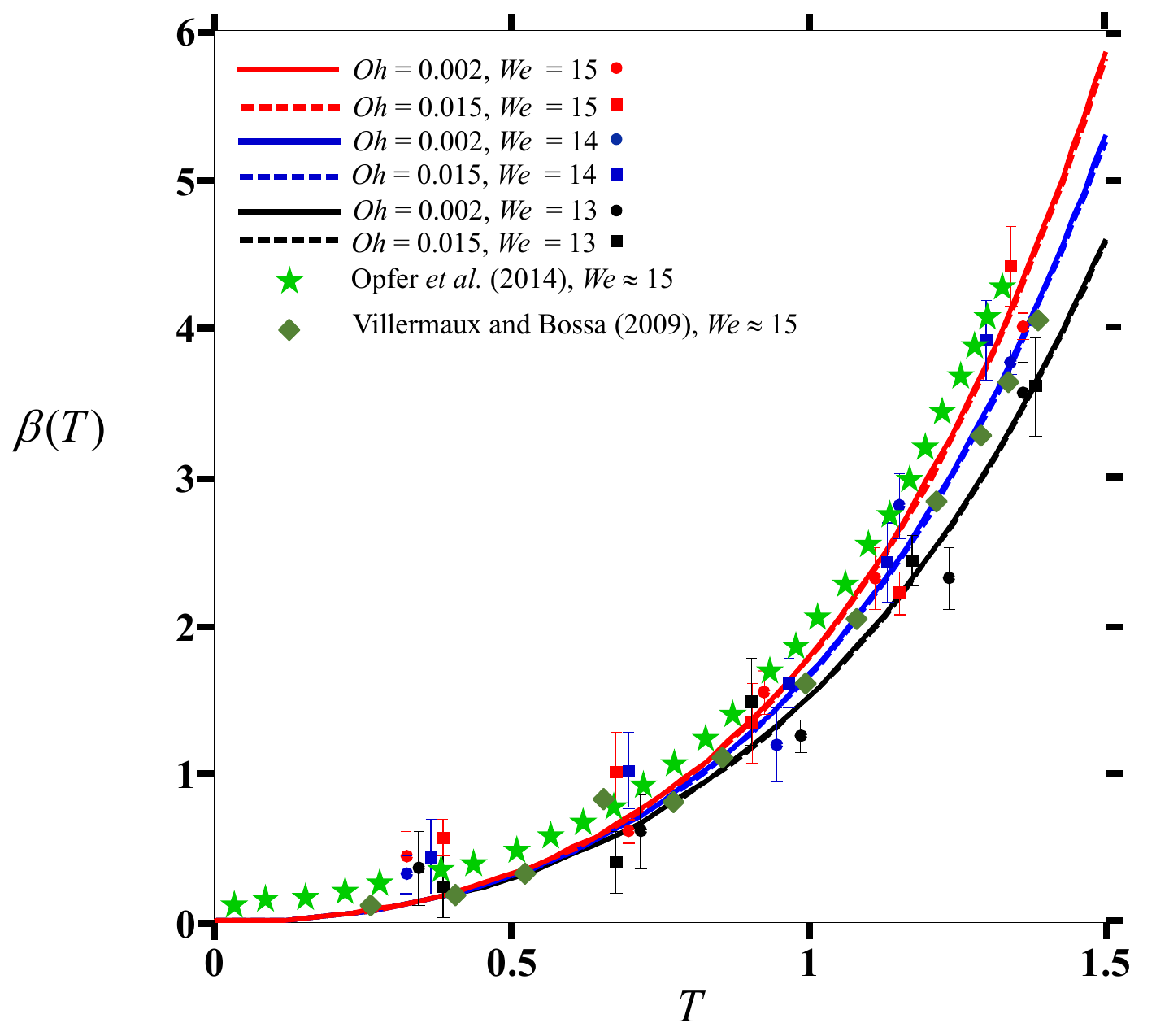}
   \vspace{0pt}
  \caption{Variation of $\beta\left(T\right)$ vs $T$ for varying $We$ and $Oh = 0.002, 0.015$. Symbols represent experimental data whereas lines represent theoretical results.}
  \label{Fig12}
\end{figure}
However, we need $\phi\left(T\right)$ for a viscous drop which can be obtained by solving \eqref{Eqn12} analytically. Since, this is not possible we circumvent these manipulations by using \eqref{Eqn23} in \eqref{Eqn27}. This is justified from Ref.\citen{Hsiang1995} experimental data where we see that for $Oh > 0.3$ the oscillatory mode ceases to exist and 7\% change in $\phi\left(T\right)$ is seen between $Oh = 0$ and $Oh = 0.3$ at $T = 2$ which roughly corresponds to the end of the bag expansion process. \eqref{Eqn28} is, thus, arrived at and includes the effect of viscosity and surface tension \footnote{The published manuscript has erroneous viscosity term which has been corrected here, also see appendix section \ref{apx3}}:

\begin{equation}\label{Eqn28}
\dfrac{d^2 \beta}{dT^2} - \dfrac{48}{We}\beta - \left[\dfrac{12Oh}{\sqrt{We}}\beta^{-1}\dfrac{d\beta}{dT} + 2\right]e^{T\sqrt{8 - \frac{96}{We}}} = 0
\end{equation}

Setting $Oh = 0$ we recover \eqref{Eqn27}. Figures \ref{Fig11} and \ref{Fig12} show \eqref{Eqn28} as $\beta \left(T\right)$ varies with $T$. For a given $We$ at a particular instant of time, $T$, the value of $\beta \left(T\right)$ is lower though not significant for a drop with higher $Oh$ in the range tested (Fig. \ref{Fig11}). At $Oh$ greater than the ones experimented here this may be more evident. This indicates that the inclusion of viscosity amounts to a delay in the bursting of the bag. Furthermore, for a fixed $Oh$, $\beta \left(T\right)$ is greater for higher $We$ at a given time (Fig. \ref{Fig12}).

\section{SUMMARY AND CONCLUSIONS}

Drop deformation dynamics are analyzed including the effects of $We$ and $Oh$. The radial extent and bag growth are quantified based on these considerations and an exponential growth in both is noted. The effect of increasing $We$ is to increase the bag growth and radial extent of the deformed drop at a given time. A small change in $We$ results in large difference in the growth extent which is especially observed at large $T$. In contrast, increasing $Oh$ dampens this growth albeit by a very small amount for the solutions considered in this work. The transition $We$ which marks the crossover from vibrational mode to bag breakup is theoretically determined and compared with existing experimental results. The derived expression shows good agreement with these observed experimental values and is an improvement on the relations obtained thus far in literature.

\section*{ACKNOWLEDGMENTS}
The authors wish to thank Dr. Daniel R. Guildenbecher for his suggestions during the preparation of the manuscript. 

\textit{\textbf{Author's note}}: This version contains corrections to the original published manuscript (Kulkarni, V., and P. E. Sojka. "Bag breakup of low viscosity drops in the presence of a continuous air jet." Physics of Fluids 26.7 (2014): 072103.) and indicated by footnotes at the corresponding places.

\section*{APPENDIX: MASS CONSERVATION EQUATION, VELOCITY FIELD INSIDE THE LIQUID DROP AND BAG GROWTH}
\appendix
\renewcommand{\thesection}{\arabic{section}} 

\section{Derivation for conservation of mass within the liquid drop \label{apx1}}  

\begin{figure}[htp!]
   \centering
   \vspace{0pt}
   \includegraphics[scale=0.40]{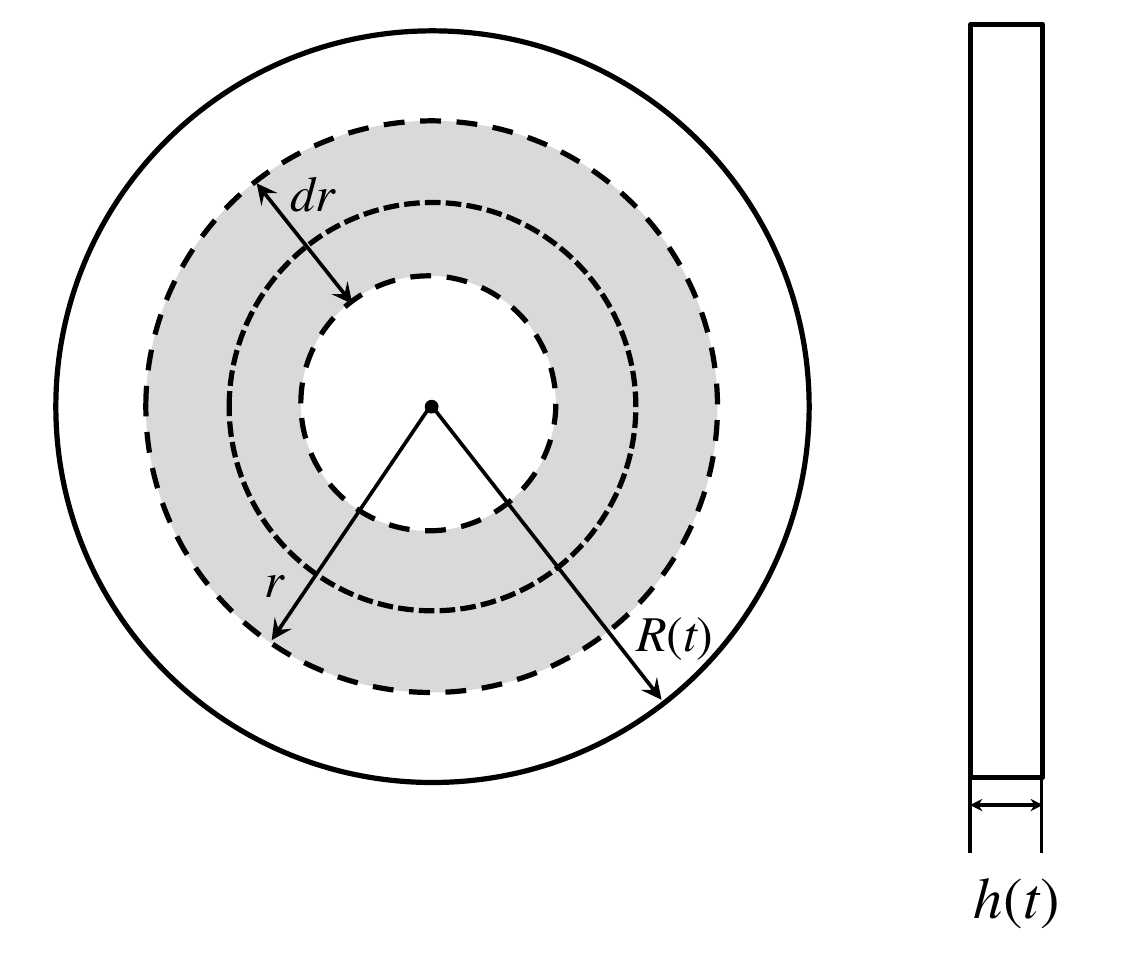}
   \vspace{0pt}
  \caption{Sketch showing two views (left: front view, right: top view) of the deformed drop with various parameters and the differential element (shaded region) considered to develop the conservation of mass equation. The dotted line within this region represents the center of the region considered.}
  \label{Fig13}
\end{figure}

Consider the deformed drop as shown in Fig. \ref{Fig13} and apply the Reynold’s Transport Theorem \eqref{A1label} to the differential mass in the figure above:
\begin{equation}\tag{A1}\label{A1label}
\dfrac{D}{Dt} \left[\;\int \displaylimits_{V_{sys}} \rho dV\;\right] = \dfrac{d}{dt} \left[\;\int \displaylimits_{CV} \rho dV\;\right] + \int \displaylimits \rho \mathbf{u}_{rel}.d\mathbf{A}
\end{equation}
The system here is the drop and its mass remains a constant thus the LHS of the above expression becomes zero and we obtain,
\begin{equation}\tag{A2}\label{A2label}
\dfrac{d}{dt} \left[\;\int \displaylimits_{CV} \rho dV\;\right] + \int \displaylimits_{CS} \rho \mathbf{u}_{rel}.d\mathbf{A} = 0
\end{equation} 
where,
\begin{equation}\tag{A3}\label{A3label}
\dfrac{d}{dt} \left[\;\int \displaylimits_{CV} \rho dV\;\right]  = \dfrac{\partial}{\partial t}\left(2 \pi \rho r h\right) dr
\end{equation}

\begin{align*}\tag{A4}\label{A4label}
\int \displaylimits_{CS}  \rho \mathbf{u}_{rel}.d\mathbf{A}  = -\left[\rho_l u_r \left(2\pi rh\right) +  \dfrac{\partial}{\partial r}\left(2 \pi \rho_l r u_r h\right) \left(-\dfrac{dr}{2}\right) \right] \\ + \left[\rho_l u_r \left(2\pi rh\right) + \dfrac{\partial}{\partial r}\left(2 \pi \rho_l r u_r h\right) \left(\dfrac{dr}{2}\right) \right] \\
= \dfrac{\partial}{\partial r}  \left(2 \pi \rho_l r u_r h \right) \qquad\qquad\qquad\qquad\qquad\;\;\;\;\;
\end{align*}

Note that, $\dfrac{dm}{dt}\Big|_{inner/outer} =  \dfrac{dm}{dt}\Big|_{middle} \pm \dfrac{\partial}{\partial r} \left(\dfrac{dm}{dt}\Big|_{middle}\right)  \left[\dfrac{dr}{2} \right]$, where, $\dfrac{dm}{dt}\Big|_{middle} = 2\pi \rho_lu_rrh$.

Plugging \eqref{A3label} and  \eqref{A4label} in  \eqref{A2label} we get,

\begin{equation}\tag{A5}\label{A5label}
r\dfrac{\partial h}{\partial t} + \dfrac{\partial}{\partial r}\left(r u_r h\right) = 0 
\end{equation}

\section{Expression for \textit{u}\textsubscript{\textit{r}}(\textit{r}, \textit{t}) \label{apx2}}  

Rewriting \eqref{A5label} we have,
\begin{equation}\tag{A6}\label{A6label}
\dfrac{\partial \left(r u_r\right)}{\partial r} = -\dfrac{r}{h}\dfrac{\partial h}{\partial t}
\end{equation}
Using global mass conservation, i.e., \eqref{Eqn7} in \eqref{A5label} transforms \eqref{A6label} to,
\begin{equation}\tag{A7}\label{A7label}
\dfrac{\partial \left(r u_r\right)}{\partial r} = -\dfrac{2r}{R}\left(\dfrac{dR}{dt}\right)
\end{equation}
Integrating \eqref{A7label} we get,
\begin{equation}\tag{A8}\label{A8label}
u_r = \dfrac{r}{R}\left(\dfrac{dR}{dt}\right) + \dfrac{f\left(t\right)}{r}
\end{equation}
At $r = 0$ the term, $\frac{f\left(t\right)}{r}$ goes to $\infty$. Since, $f\left(t\right)$ is an arbitrary function of $t$ this can only be true if it takes the value zero. Thus, we obtain \eqref{Eqn8}.
\vspace{15pt}

\section{Differential equation for bag growth: inviscid and viscous case \label{apx3}} 

In this section we provide details of the derivation corresponding to \eqref{Eqn27} and \eqref{Eqn28}. To begin, we consider the force balance at the bag tip (of unit area) as outlined in Ref \citen{Villermaux2009},
\begin{equation}\tag{A9}\label{A9label}
\rho_lh\left(t\right)\dfrac{d^2\alpha}{dt^2} =  \rho_aU^2 + \dfrac{4\sigma}{r_c} + 2\mu_l\dfrac{\partial u_c}{\partial r_c} 
\end{equation}

Following the Rayleigh-Plesset equation for spherical geometries we may write the expression for the liquid flow velocity within the expanding bag as, $u_{r_c } = \frac{r_c^2}{r^2}\dfrac{dr_c}{dt}$. Using the simplification, $R^2\left(t\right) > \alpha\left(t\right)$ we arrive the concise form of \eqref{Eqn26}, $r_c\left(t\right) = R^2\left(t\right)/2\alpha\left(t\right)$. Further, noting that in the bag growth stage, $\frac{dR\left(t\right)}{dt} \approx 0$ we can write the viscous term of \eqref{A9label} in terms of known variables thereby transforming \eqref{A9label} to,

\begin{equation}\tag{A10}\label{A10label}
\rho_lh\left(t\right)\dfrac{d^2\alpha}{dt^2} =  \rho_aU^2 + \dfrac{4\sigma}{r_c} + 2\mu_l\left(\dfrac{2}{\alpha}\dfrac{d\alpha}{dt}\right)
\end{equation}

Substituting the non-dimensional variables, $\beta\left(t\right) = \dfrac{\alpha \left(t\right)}{d_0/2}$, $H\left(t\right) = \dfrac{h\left(t\right)}{d_0/2}$, $\phi\left(t\right) = \dfrac{R\left(t\right)}{d_0/2}$ and $T = t/\frac{d_0}{U}\sqrt{\frac{\rho_l}{\rho_a}}$ we arrive \eqref{Eqn27} and \eqref{Eqn28} which represent the inviscid and viscous forms of the bag growth evolution differential equations, respectively.

\bibliography{aipsamp}

\end{document}